\documentclass[prl,superscriptaddress,twocolumn,aps]{revtex4}

\usepackage{psfig}

\begin{document}

\psfigurepath{.:figures}

\title{Novel Coexistence of 
Superconductivity with Two Distinct Magnetic Orders
}
\author{A.D. Christianson}
\altaffiliation{This work was part of ADC's Ph.D.\ research project, supervised by WB and AHL at LANL.}
\affiliation{Los Alamos National Laboratory, Los Alamos NM 87545}
\affiliation{Colorado State University, Fort Collins CO 80523}
\author{A. Llobet}
\author{Wei Bao}
\affiliation{Los Alamos National Laboratory, Los Alamos NM 87545}
\author{J.S. Gardner}
\altaffiliation[Current address: ]{Physics Department, Brookhaven National Laboratory, Upton, New York 11973}
\affiliation{NRC Canada, NPMR, Chalk River Laboratories, Chalk
River, Ontario KOJ IJO, Canada}
\affiliation{NIST Center for Neutron Research, National Institute of Standards and Technology, Gaithersburg, MD 20899}
\author{I.P. Swainson}
\affiliation{NRC Canada, NPMR, Chalk River Laboratories, Chalk
River, Ontario KOJ IJO, Canada}
\author{J.W. Lynn}
\affiliation{NIST Center for Neutron Research, National Institute of Standards and Technology, Gaithersburg, MD 20899}
\author{J.-M. Mignot}
\affiliation{Laboratoire L\'{e}on Brillouin, CEA-CNRS, CEA/Saclay, 91191 Gif-sur-Yvette, France}
\author{K. Prokes}
\affiliation{BENSC, Hahn-Meitner-Institut, Glienickerstrass 100, D-14109, Berlin, Germany}
\author{P.G. Pagliuso}
\altaffiliation[Current address: ]{Instituto de F\'{i}sica ``Gleb Wathagin'', UNICAMP, 13083-970, Campinas, Brasil}
\author{N.O. Moreno}
\author{J.L. Sarrao}
\author{J.D. Thompson}
\author{A.H. Lacerda}
\affiliation{Los Alamos National Laboratory, Los Alamos NM 87545}

\date{\today}

\begin{abstract}
The heavy fermion CeRh$_{1-x}$Ir$_x$In$_5$ system exhibits properties that range from an incommensurate antiferromagnet for small $x$ to an exotic superconductor on the Ir-rich end of the phase diagram.  At intermediate $x$ where antiferromagnetism coexists with superconductivity, 
two types of magnetic order are observed: the {\em incommensurate} one of 
CeRhIn$_5$ and a new, {\em commensurate} antiferromagnetism that orders separately.  The coexistence of $f$-electron superconductivity with two distinct $f$-electron magnetic orders is unique among unconventional superconductors, adding a new variety to the usual coexistence found in magnetic superconductors.
\end{abstract}

\pacs{??}

\maketitle

Magnetism and superconductivity are two major cooperative
phenomena in condensed matter, and the relationship between them has 
been studied extensively. Conventional superconductivity 
in phonon mediated $s$-wave materials is susceptible to the Cooper-pair 
breaking by magnetic scattering\cite{mswave,eswave}.
In cases where superconductivity and magnetic order coexist,
such as rare-earth based molybdenum chalcogenides, rhodium 
borides and borocarbides, the superconducting
$d$ electrons and localized $f$ electrons are weakly 
coupled, and the competition between superconductivity and magnetic 
order is understood\cite{eswave,bswave} in terms of the theory of
Abrikosov and Gorkov\cite{mswave}. In contrast, 
strong magnetic fluctuations have been observed in 
heavy fermion, cuprate and ruthenate 
superconductors\cite{agold,ru_ins,upd_gap,upd_gapa}, and
magnetic excitations have been proposed to mediate the Cooper pairing 
in these unconventional 
superconductors\cite{rmp_upt3,msrv,msrv1,msrv3,pmdp,pwave}.
Some of the U-based heavy fermion compounds form an interesting
subset of these superconductors where
superconductivity develops out of a magnetically ordered state. 
For UPd$_2$Al$_3$\cite{upd_sc,upd_af} the magnetic state is 
a commensurate antiferromagnet; for 
UNi$_2$Al$_3$\cite{uni_sc,uni_af} an incommensurate antiferromagnet; 
for UGe$_2$\cite{uge2} and URhGe\cite{urhge} 
a ferromagnet. For UPt$_3$\cite{upt3_afsc}, three distinct 
superconducting phases\cite{sc_upt3,upt3_sca,upt3_scc} 
coexist with a commensurate short-range antiferromagnetic order.     
In analog to superfluid $^3$He, intriguing coupling 
between the magnetic and superconducting order parameters has been
proposed\cite{rmp_upt3}.

\begin{figure}
\psfig{file=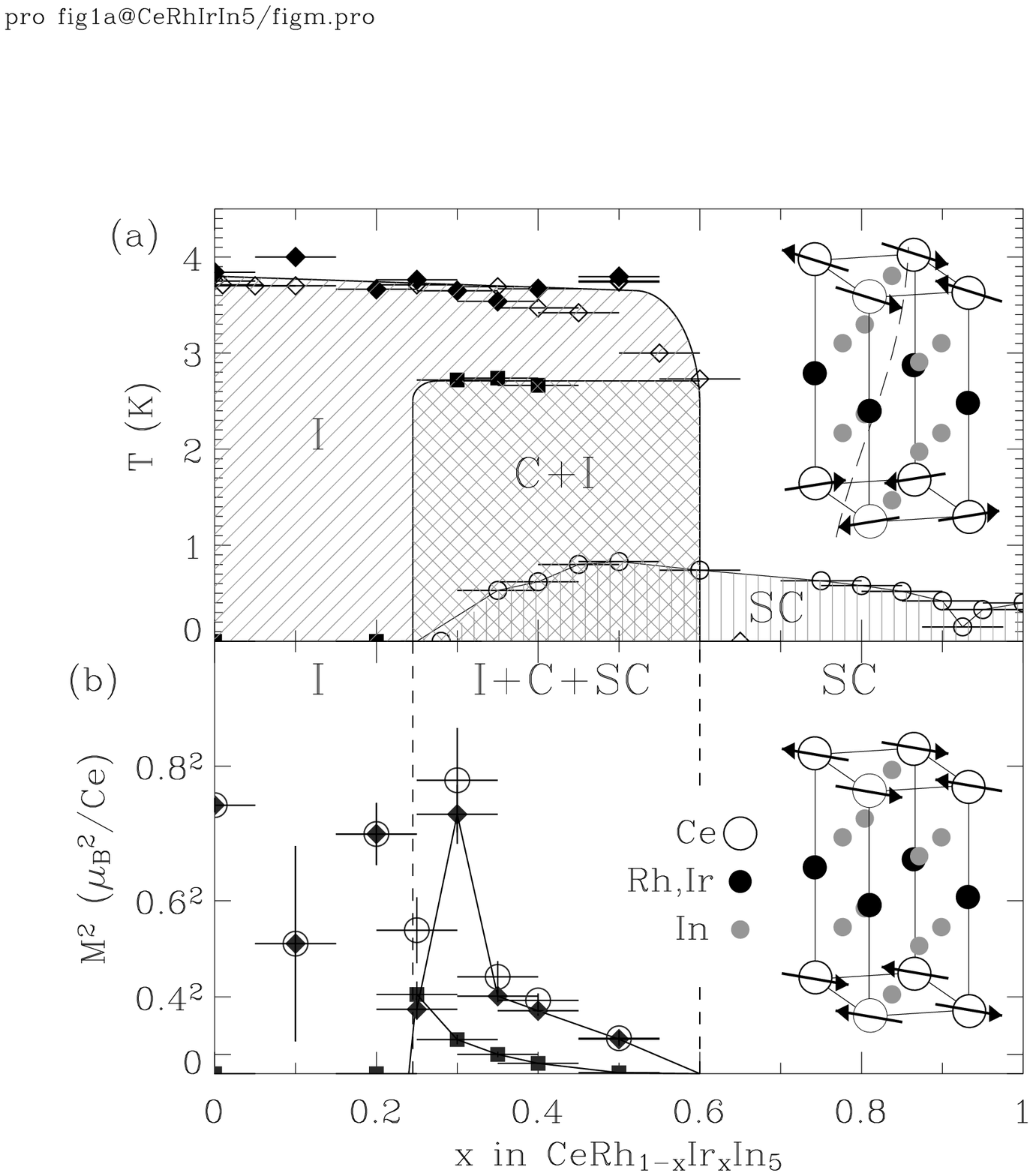,width=.9\columnwidth,angle=0,clip=}
\vskip -11 mm
\caption{(a) Temperature-composition phase diagram for 
CeRh$_{1-x}$Ir$_x$In$_5$. There are three distinct long-range orders:
superconducting (SC), incommensurate antiferromagnetic (I), 
and commensurate antiferromagnetic (C).
The diamonds represent the N\'{e}el temperature for the I phase, 
squares for the C phase, and circles for $T_C$ of the SC phase. 
The open symbols are from previous work\cite{pgpRhIr}, and
the closed symbols from this work. 
(b) Diamonds and squares represent squared magnetic moments at 1.9 K 
for the I and C phases,
respectively. The circles represent the square of the total magnetic moment per Ce in Eq.~(\ref{totm}).
The solid lines are guides to the eye. The dashed lines delimit the
I, I+C+SC and SC phases.
Magnetic structures of the I and C phases are shown as insets to (a) and (b), respectively.
}
\label{fig1}
\end{figure}

Recently a new family of Ce-based heavy fermion materials has been
discovered, which sets a record superconducting
transition temperature of $T_C$=2.3 K for heavy fermion 
materials\cite{hegger,joeIr,joe218}. Coexistence of
magnetic order and superconductivity is observed in a wide
composition range in CeRh$_{1-x}$Ir$_x$In$_5$\cite{pgpRhIr} 
[see Fig.~\ref{fig1}(a)]; previously superconductivity was found only
in heavy fermion materials of the highest purity\cite{uge2}.
At one end of the CeRh$_{1-x}$Ir$_x$In$_5$ series, CeIrIn$_5$ is a 
superconductor below $T_C$=0.4 K\cite{joeIr}. At the other end,
CeRhIn$_5$ orders magnetically below $T_{Ni}$=3.8 K in an 
antiferromagnetic spiral structure with an ordering
wave vector ${\bf q}_i$=$(\frac{1}{2},\frac{1}{2}, \pm\delta)$, 
$\delta$=0.297\cite{bao00a}, that is
incommensurate with the tetragonal crystal lattice\cite{nstru}.
In the overlapping region of superconducting and antiferromagnetic phases,
$0.25 \lesssim x \lesssim 0.6$, we find unexpectedly an additional commensurate antiferromagnetic 
order that develops separately below $T_{Nc}$=2.7 K. 
This makes Ce(Rh,Ir)In$_5$
unique among unconventional superconductors in that the ground state 
exhibits the coexistence of the superconducting order parameter with
two distinct magnetic order parameters.

Single crystals of CeRh$_{1-x}$Ir$_x$In$_5$ were grown 
from an In flux with appropriate ratio of Rh and Ir starting 
materials\cite{pgpRhIr}. Lattice parameters  
follow Vegard's law and the sample composition $x$ is estimated to be
within $\pm 0.05$ of the nominal composition\cite{pgpRhIr}.
Samples used in this work were cut to a thickness $\sim$1.3 mm 
to minimize problems due to the high neutron absorption of Rh, Ir
and In.  The search for and collection of magnetic Bragg neutron 
diffraction peaks in CeRh$_{1-x}$Ir$_x$In$_5$ ($x=$0.1, 0.2, 0.25,
0.3, 0.35, 0.4, and 0.5) were performed at the thermal triple-axis 
spectrometers C5 and N5 of Chalk River Laboratories, and BT2 and BT7 of NIST.  
Neutrons with incident energy E$_i$=35 meV were selected
using the (113) reflection of a Ge crystal, the (002) of a Be
crystal, or the (002) of a pyrolytic graphite monochromator. 
At this energy, the neutron penetration length is longer than the sample
thickness. Polarized neutron diffraction experiments were performed
at BT2 of NIST with E$_i$=14.7 meV on a $x$=0.3 sample 
to verify magnetic signals and to determine magnetic moment orientation. 
Pyrolytic graphite filters were employed when appropriate
to remove higher order neutrons. The sample temperature was controlled 
by a top-loading pumped $^4$He cryostat at both Chalk River and NIST. 
Additional measurements of magnetic order parameters
were performed at the triple-axis spectrometer 4F2 of LLB-Saclay using a pumped $^4$He cryostat, 
and at
the E4 diffractometer of BENSC using a dilution refrigerator.

In a previous heat capacity study\cite{pgpRhIr}, a phase transition was 
observed in the range of 3.8 to 2.7 K for $0\le x \lesssim 0.6$, and was attributed to an 
antiferromagnetic transition as in CeRhIn$_5$\cite{hegger} 
[open diamonds in Fig.~\ref{fig1}(a)].
This is confirmed by the results of this neutron diffraction work which
show that all samples
($0.1 \le x \le 0.5$) have magnetic Bragg peaks characterized by the
same incommensurate wave vector as in CeRhIn$_5$\cite{bao00a}.
The N\'{e}el temperature, $T_{Ni}$, determined from the temperature
variation of magnetic Bragg peaks is shown as filled diamonds
in Fig.~\ref{fig1}(a).

Thermodynamic and transport measurements uncover 
a superconducting state below 1 K in a wide composition range,
$0.25 \lesssim x \le 1$\cite{pgpRhIr},
[open circles in Fig.~\ref{fig1}(a)].
In $0.25 \lesssim x \lesssim 0.6$ where the superconducting and 
incommensurate antiferromagnetic phases coexist,
we find a second magnetic order below 2.7 K with
a commensurate antiferromagnetic ordering wave vector 
${\bf q}_c$=$(\frac{1}{2},\frac{1}{2},\frac{1}{2})$
[filled squares in Fig.~\ref{fig1}].

Table I lists integrated intensities of magnetic Bragg peaks
\begin{table}
\caption{Magnetic Bragg intensity, $\sigma_{obs}$,
defined in Eq.~(\ref{eq_cs}), observed at
1.9~K in units of $10^{-3}$ barns per CeRh$_{0.7}$Ir$_{0.3}$In$_5$.
The theoretical intensity, $\sigma_{cal}$, is calculated using Eq.~(\ref{eq_ic}) and (\ref{eq_cm}) for the incommensurate moment
$M_i=0.73\mu_B$/Ce and the commensurate moment $M_c=0.27\mu_B$/Ce,
respectively. 
}
\label{mlist}
\begin{ruledtabular}
\begin{tabular}{ccc|ccc}
${\bf q}$ & $\sigma_{obs}$ & $\sigma_{cal}$ &
${\bf q}$ & $\sigma_{obs}$ & $\sigma_{cal}$\\
\hline
($\frac{1}{2}$ $\frac{1}{2}$ $\delta$  ) &   9.0(5) &   9.3 &
($\frac{1}{2}$ $\frac{1}{2}$ 1-$\delta$  ) &  10.1(3) &  10.8 \\
($\frac{1}{2}$ $\frac{1}{2}$ 1+$\delta$  ) &   9.6(3) &  12.1 &
($\frac{1}{2}$ $\frac{1}{2}$ 2-$\delta$  ) &  9.9(3) &  12.0 \\
($\frac{1}{2}$ $\frac{1}{2}$ 2+$\delta$  ) &  9.0(3) &  10.8 &
($\frac{1}{2}$ $\frac{1}{2}$ 3-$\delta$  ) &  8.2(4) &   9.6 \\
($\frac{1}{2}$ $\frac{1}{2}$ 3+$\delta$  ) &  5.7(3) &   7.7 &
($\frac{1}{2}$ $\frac{1}{2}$ 4-$\delta$  ) &  5.7(3) &   6.5 \\
($\frac{1}{2}$ $\frac{1}{2}$ 4+$\delta$  ) &  4.3(3) &   4.9 &
($\frac{1}{2}$ $\frac{1}{2}$ 5-$\delta$  ) &  3.9(3) &   3.9 \\
($\frac{1}{2}$ $\frac{1}{2}$ 5+$\delta$  ) &  2.9(3) &   2.7 &
($\frac{1}{2}$ $\frac{1}{2}$ 6-$\delta$  ) &  2.7(5) &   2.1 \\
($\frac{3}{2}$ $\frac{3}{2}$ $\delta$  ) &   4.3(3) &   4.2 &
($\frac{3}{2}$ $\frac{3}{2}$ 1-$\delta$  ) &  3.8(4) &   4.2 \\
($\frac{3}{2}$ $\frac{3}{2}$ 1+$\delta$  ) &  5.4(15) & 4.2 &
($\frac{3}{2}$ $\frac{3}{2}$ 2-$\delta$  ) &  5.0(4) &   4.1 \\
($\frac{3}{2}$ $\frac{3}{2}$ 2+$\delta$  ) &  3.4(3) &   3.9 &
($\frac{3}{2}$ $\frac{3}{2}$ 3-$\delta$  ) &  2.9(3) &   3.6 \\
($\frac{3}{2}$ $\frac{3}{2}$ 3+$\delta$  ) &  2.5(4) &   3.1 &
($\frac{1}{2}$ $\frac{1}{2}$ $\frac{1}{2}$) &  2.3(2) &   2.6 \\
($\frac{1}{2}$ $\frac{1}{2}$ $\frac{3}{2}$) &   2.3(4) &   3.2 &
($\frac{1}{2}$ $\frac{1}{2}$ $\frac{5}{2}$) &   1.9(7) &   2.7 \\
($\frac{1}{2}$ $\frac{1}{2}$ $\frac{9}{2}$) &   0.8(2) &   1.1 &
($\frac{3}{2}$ $\frac{3}{2}$ $\frac{1}{2}$) &   1.3(2) &   1.1 \\
($\frac{3}{2}$ $\frac{3}{2}$ $\frac{3}{2}$) &   1.2(2) &   1.1 &
($\frac{3}{2}$ $\frac{3}{2}$ $\frac{5}{2}$) &   0.7(2) &   1.0 \\
\end{tabular}
\end{ruledtabular}
\end{table}
at 1.9 K for $x=0.3$
of both commensurate and incommensurate types, collected by rocking
scans in the two-axis mode and normalized to structural Bragg peaks
(002), (003), (004), (006), (111), (112), (113), (220), (221) and (222) 
to yield magnetic neutron diffraction cross-sections in absolute units,
$\sigma({\bf q})=I({\bf q})\sin(2\theta)$, where $2\theta$ is the scattering angle.
In such units, the magnetic cross section is\cite{neut_squire}
\begin{equation}
\sigma({\bf q})=\left(\frac{\gamma r_0}{2}\right)^2
        M_{i,c}^2 \left|f(q)\right|^2
        \sum_{\mu,\nu}(\delta_{\mu\nu}
        -\widehat{\rm q}_{\mu}\widehat{\rm q}_{\nu})
        {\cal F}^*_{\mu}({\bf q}){\cal F}_{\nu}({\bf q}),
\label{eq_cs}
\end{equation}
where $(\gamma r_0/2)^2=0.07265$~barns/$\mu_B^2$, $M_{i,c}$ is
the staggered moment of the Ce ion in either the incommensurate or
commensurate antiferromagnetic structure, $f(q)$ the magnetic
form factor which could be different for the two type 
antiferromagnetic orders, $\widehat{\bf q}$ the unit vector of ${\bf q}$,
and ${\cal F}_{\mu}({\bf q})$ the $\mu$th
Cartesian component of the magnetic structure factor per molecular formula.

The incommensurate magnetic structure for CeRhIn$_5$ has been 
determined [inset in Fig.~\ref{fig1}(a)]
and the magnetic cross section, Eq.~(\ref{eq_cs}), is
reduced to\cite{bao00a}
\begin{equation}
\sigma^i({\bf q})=\frac{1}{4}\left(\frac{\gamma r_0}{2}\right)^2
        M_i^2 \left|f(q)\right|^2
        \left(1+|\widehat{\bf q}\cdot \widehat{\bf c}|^2
        \right),
\label{eq_ic}
\end{equation}
where $\widehat{\bf c}$ is the unit vector of the c-axis.

For the commensurate magnetic component, 
the summation in Eq.~(\ref{eq_cs}) is reduced to
$(1-|\widehat{\bf q}\cdot \widehat{\bf M}_c|^2 )
\left|{\cal F}({\bf q})\right|^2$,
where $\widehat{\bf M}_c$ is the unit vector of the magnetic moment,
and $\left|{\cal F}({\bf q})\right|^2=1$. Polarized neutron diffraction
measurements at $(\frac{1}{2},\frac{1}{2},\frac{1}{2})$ and
$(\frac{1}{2},\frac{1}{2},\delta)$, with a horizontal and vertical 
magnetic guide field respectively, indicate that the magnetic moment 
in the commensurate antiferromagnetic order, like in the incommensurate 
order\cite{bao00a}, lies in the tetragonal basal plane. 
The commensurate magnetic structure is 
depicted in the inset in Fig.~\ref{fig1}(b). There are in general 
8 symmetry-related $\widehat{\bf M}_c$ domains in the sample and the easy axis
within the basal plane cannot be determined in 
a multidomain sample due to the tetragonal symmetry. 
Assuming equal populations among these domains during our unpolarized
neutron diffraction experiment, the magnetic polarization factor 
averages to\cite{bao01a}
$\langle 1-|\widehat{\bf q}\cdot \widehat{\bf M}_c|^2 \rangle
=\left(1+|\widehat{\bf q}\cdot \widehat{\bf c}|^2\right)/2$.
Therefore, the magnetic cross section for the commensurate 
antiferromagnetic order is 
\begin{equation}
\sigma^c({\bf q})=\frac{1}{2}\left(\frac{\gamma r_0}{2}\right)^2
        M_c^2 \left|f(q)\right|^2
        \left(1+|\widehat{\bf q}\cdot \widehat{\bf c}|^2
        \right).
\label{eq_cm}
\end{equation}
Applying Eq.~(\ref{eq_ic}) and (\ref{eq_cm}) to the data in Table I,
a least-square fit yields staggered magnetic moments 
$M_i$=$0.73\pm 0.25\mu_B$ and $M_c$=$0.27\pm 0.10 \mu_B$ 
per CeRh$_{0.7}$Ir$_{0.3}$In$_5$ at 1.9~K.

\begin{figure}
\psfig{file=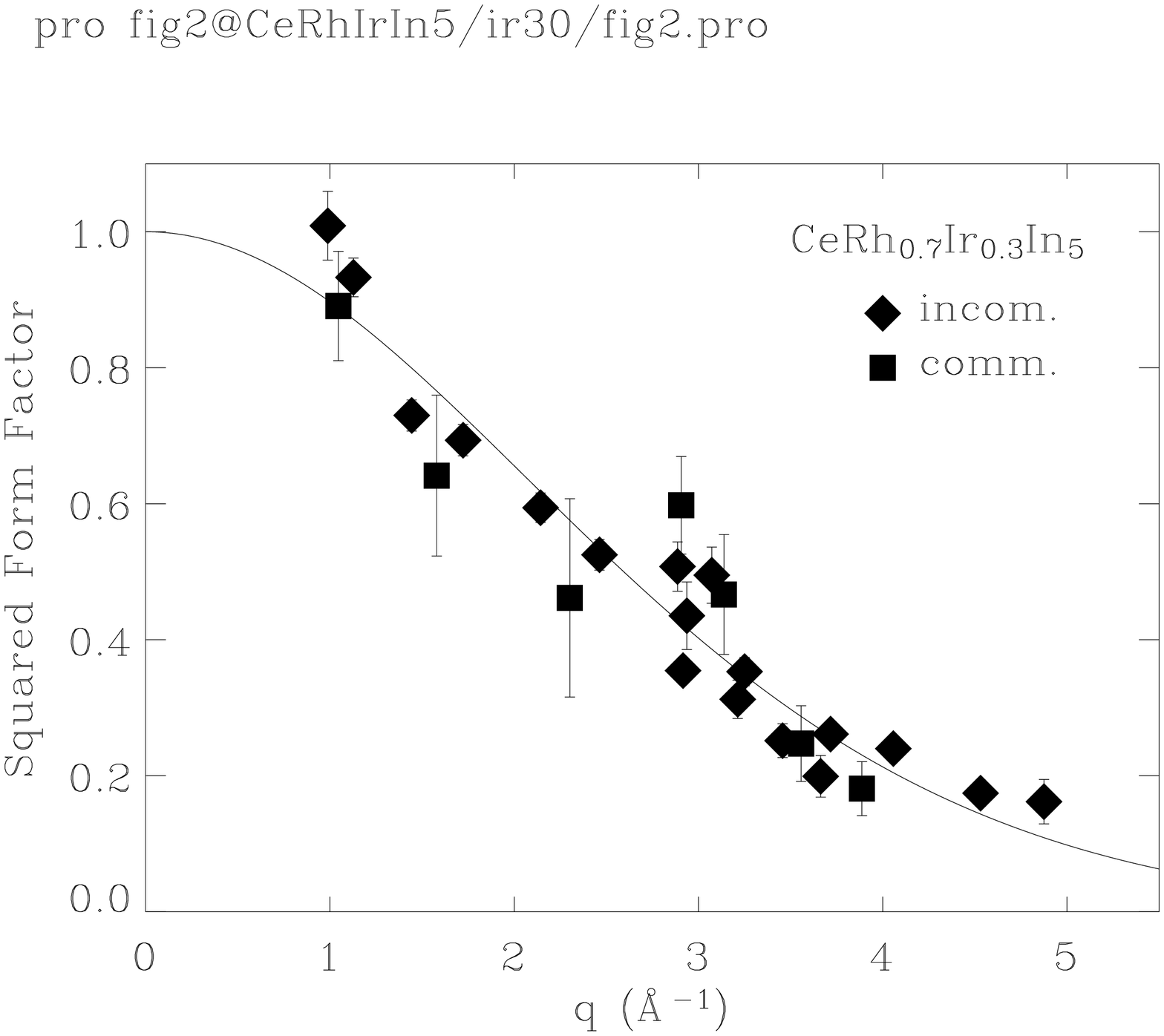,width=.8\columnwidth,angle=90,clip=}
\vskip -4 mm
\caption{
Both the commensurate (squares) and incommensurate (diamonds) magnetic
Bragg peaks follow the magnetic form factor of Ce$^{3+}$ (4f$^1$) ion.
}
\label{fig2}
\end{figure}

Measured magnetic intensities for CeRh$_{0.7}$Ir$_{0.3}$In$_5$
in Table I are plotted in Fig.~\ref{fig2} 
as a function of $|{\bf q}|$ in scaled quantities $4\sigma^i({\bf q})/\left[ (\gamma r_0/2 )^2
        M_i^2 
         (1+|\widehat{\bf q}\cdot \widehat{\bf c}|^2
         )\right]$
and $2\sigma^c({\bf q})/\left[ (\gamma r_0/2 )^2
        M_c^2 
         (1+|\widehat{\bf q}\cdot \widehat{\bf c}|^2
         )\right]$, 
which equal the squared magnetic form factor, 
$\left|f(q)\right|^2$, [Eq.~(\ref{eq_ic}) and (\ref{eq_cm})].
The solid line is $\left|f(q)\right|^2$ for Ce$^{3+}$
ion\cite{formf_ce}.
The fact that the measured data for both 
commensurate (squares) and incommensurate (diamonds) orders
are well described by the solid curve indicates
not only the correctness of our magnetic models in 
Eq.~(\ref{eq_ic}) and (\ref{eq_cm}) for CeRh$_{0.7}$Ir$_{0.3}$In$_5$, 
but also that the Ce$^{3+}$ ($4f^1$) ions are responsible for both 
commensurate and incommensurate antiferromagnetic orders.
This is consistent with the thermodynamic study\cite{pgpRhIr} where 
entropy reaches the same value at $\sim$6 K for
all samples from $x$=0 to $x$=1, suggesting that the same $f$ electrons
are responsible for heavy fermion formation, antiferromagnetic orders,
and superconductivity in CeRh$_{1-x}$Ir$_{x}$In$_{5}$.
Division of the $f$ spectral weight was probed using neutron scattering in CeRhIn$_5$\cite{bao01b}
and was further investigated in terms of a two-fluid 
model for this family of heavy fermion materials\cite{tf_bang,tf_nick,tf_sat}.

Magnetic moments for other compositions were measured in 
a similar fashion, and 
are shown as squares and diamonds for commensurate and incommensurate structures, respectively, 
in Fig.~\ref{fig1}(b). The average of the squared
total magnetic moments per Ce,
\begin{equation}
 M^2\equiv \langle \left({\bf M}_i+{\bf M}_c\right)^2\rangle 
=M_i^2+M_c^2, 
\label{totm}
\end{equation}
is also shown in Fig.~\ref{fig1}(b) as circles. A large staggered magnetic
moment of 0.85$\mu_B$/U\cite{upd_af} coexisting with superconductivity 
in UPd$_2$Al$_3$ has been considered a puzzling anomaly\cite{rmp_upt3}, 
while UPt$_3$\cite{upt3_afsc} has
a tiny moment of 0.02 $\mu_B$/U, and the moment in 
UNi$_2$Al$_3$, 0.24 $\mu_B$/U\cite{uni_af}, is also quite small.  
Here in the coexistence composition region of
Ce(Rh,Ir)In$_5$, $M$ ranges
from 0.27 to 0.78 $\mu_B$/Ce, bridging the gap between
UNi$_2$Al$_3$ and UPd$_2$Al$_3$.

\begin{figure}
\psfig{file=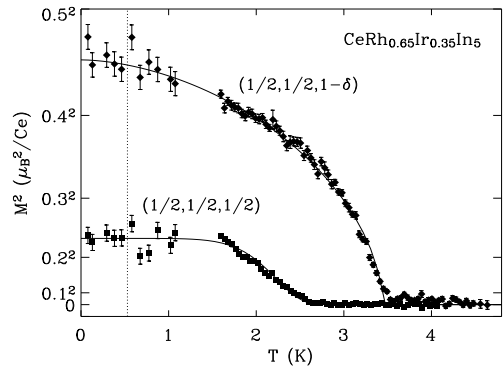,width=.85\columnwidth,angle=90,clip=}
\caption{Squared magnetic order parameters of the incommensurate (diamonds)
and commensurate (squares) antiferromagnetic transitions. The dotted line
indicates the critical temperature $T_C\approx 0.5$ K of the superconducting transition. The solid lines are guides to the eye.
}
\label{fig3}
\end{figure}

In Fig.~\ref{fig3}, the temperature dependences of the Bragg peak intensities of 
($\frac{1}{2},\frac{1}{2},1$$-$$\delta$) and
$(\frac{1}{2},\frac{1}{2},\frac{1}{2})$ are presented as the squared order
parameters $M_i^2$ and $M_c^2$ of the incommensurate and commensurate antiferromagnetic transitions,
respectively. Data below 1.2 K were taken at BENSC using a dilution 
refrigerator. There is no thermal hysteresis for either
magnetic phase transition. When the commensurate phase
transition occurs at 2.7 K, there is no abrupt change in the
incommensurate order parameter. Similarly, at the superconducting 
transition $T_C\approx 0.5$ K (dotted line),
no apparent anomaly occurs in either magnetic
order parameter. This is analogous
to the magnetic order parameter for the heavy fermion superconductors
UPd$_2$Al$_3$\cite{upd_aftc} and UPt$_3$ in the A and C phases,
but different from that for UPt$_3$ in the superconducting B 
phase\cite{upt3_afsc,rmp_upt3}.

Note in Fig.~\ref{fig1}(a) that the N\'{e}el temperatures $T_{Ni}$ and 
$T_{Nc}$ both are nearly constant when they are non-zero. 
This might suggest a phase separation scenario where only a volume 
fraction of $M^2(x)/M^2(0)$ is magnetically ordered
and the remaining is superconducting.
However, at $x=0.5$, this would imply $M^2(x)/M^2(0)=0.08$,
which is not consistent with a magnetic 
volume fraction of at least 0.85 determined from a $\mu$SR study\cite{uSR_gdm}.

At the quantum critical point (QCP) of the
superconducting phase near $x=0.25$, 
the commensurate antiferromagnetic order appears
with $M_c$ jumping from zero
[Fig.~\ref{fig1}(b)]. This suggests a change in electronic structure
at $x\approx 0.25$ which produces a strong enough peak at 
${\bf q}=(\frac{1}{2},\frac{1}{2},\frac{1}{2})$
in the RKKY interaction to induce the new magnetic order. 
Several sheets of the Fermi 
surface of differently enhanced masses have been observed in the de Haas-van Alphen measurements of
CeRhIn$_5$ and CeIrIn$_5$\cite{haga,hall_Rh} and the Fermi surface topology 
is determined mostly by non-$f$ electrons\cite{a_RhLa,ldau}.
The concurrence of the superconductivity and commensurate antiferromagnetic order observed here, therefore, suggests that the responsible Fermi surface sheet for 
the superconductivity may be the one close to nesting at 
${\bf q}=(\frac{1}{2},\frac{1}{2},\frac{1}{2})$.

Both $M_i$ and $M_c$ approach zero at a second QCP of the antiferromagnetism
near $x\approx 0.6$, while the N\'{e}el temperatures are 
insensitive to $x$
(Fig.~\ref{fig1}). This behavior
is similar to UPt$_3$ under pressure\cite{upt3_smh}, suggesting that
both belong to a distinct type of QCP. Recall that
$T_N\propto J_{RKKY}M^2$, where both the RKKY interaction $J_{RKKY}$ 
and the saturated magnetic moment $M$ are controlled
by the Kondo interaction. In both of these systems the divergence 
of $J_{RKKY}$ and the reduction of $M$ appear to obey
$J_{RKKY}\propto M^{-2}$ as this QCP is approached. Theoretical investigation of the physical process
maintaining the delicate proportional relation near this type of QCP is
warranted.

In summary, the antiferromagnetic phase of Ce(Rh,Ir)In$_5$ 
below $\sim$3.8 K is characterized by the 
incommensurate antiferromagnetic spiral of wave vector 
$(\frac{1}{2},\frac{1}{2}, \pm\delta)$. In the coexistence 
composition region of superconductivity and antiferromagnetism,
an additional phase transition to a commensurate antiferromagnet 
characterized by $(\frac{1}{2},\frac{1}{2},\frac{1}{2} )$
is discovered below 2.7 K. The same $f$ electron at each site, hybridizing
with other conduction electrons, is responsible for both the superconductivity and the commensurate and incommensurate antiferromagnetic orders.
It is likely that the energy band(s) with Fermi surface nesting 
near the $(\frac{1}{2},\frac{1}{2},\frac{1}{2})$ is responsible for the
heavy fermion superconductivity.
The novel coexistence of three different
types of cooperative ordered states adds a new 
variety to the rich phenomena relating to the interplay between 
magnetism and superconductivity.

We would like to thank 
G.D.\ Morris, 
R.H.\ Heffner, 
C.M.\ Varma, Q.\ Si, S.A.\ Trugman, D.\ Pines, Ar.\ Abanov, 
Y.\ Bang, 
P.\ Dai and S.\ Kern
for useful discussions; A.\ Cull for help at CRL, and P. Smeibidl and
S. Gerischer at HMI.
Work at Los Alamos was performed under the auspices of the US 
Department of Energy.


\end{document}